\documentclass[prb,twocolumn,preprintnumbers,amsmath,amssymb]{revtex4}

\usepackage{graphicx}
\usepackage{dcolumn}
\usepackage{bm}
\usepackage{amsmath}
\usepackage{amssymb}
\usepackage{amsthm}
\usepackage{amsfonts}
\usepackage{enumerate}
\usepackage{latexsym}

\begin{document}

\title {Semimetal with both Rarita-Schwinger-Weyl and Weyl excitations }

\author{ Long Liang$^{1,2}$ and Yue Yu$^{1,3}$}

\affiliation {1. Department of Physics, Center for Field
Theory and Particle Physics, State Key Laboratory of Surface Physics and Collaborative Innovation Center of
 Advanced Microstructures, Fudan University, Shanghai 200433,
China \\2. COMP Centre of Excellence, Department of Applied Physics, School of Science, 
Aalto University, FI-00076 Aalto, Finland\\
3. State Key Laboratory of Theoretical Physics, Institute of
Theoretical Physics, Chinese Academy of Sciences, P.O. Box 2735,
Beijing 100190, China 
}

\begin{abstract}
 A relativistic spinor with spin 3/2 is historically called Rarita-Schwinger spinor. The right- and left-handed chiral degrees of freedom for the massless Rarita-Schwinger spinor are independent and are thought of as the left- and right-Weyl fermion with helicity $\pm3/2$. We study  three orbital spin-1/2 Weyl semimetals in the strong spin-orbital coupling limit with time reversal symmetry breaking. We find that in this limit the systems can be a $J_{eff}=1/2$ Weyl semimetal or a $J_{eff}=3/2$ semimetal, depending on the Fermi level position. The latter near Weyl points includes both degrees of freedom of Rarita-Schwinger-Weyl  and Weyl's. A non-local potential separates the Weyl and Rarita-Schwinger-Weyl degrees of freedom and a relativistic Rarita-Schwinger-Weyl semimetal emerges.  This recipe can be generalized  to mulit-Weyl semimetal and Weyl fermions with pairing interaction and obtain high monopole charges.
 Similarly, a spatial inversion breaking Raita-Schwinger-Weyl semimetal may also emerge.
\end{abstract}


\maketitle

\section{introductions}

 In 1936, eight years after derived his famous relativistic electron equations of motion, Dirac generalized these equations to higher spin relativistic particles \cite{dirac}.  The first important example was the  recovering of Maxwell equations.
The next simplest particle except the electron and the photon is of spin-3/2 and obeys so-called Rarita-Schwinger (RS)  equations\cite{RS}. This fermonic field later played an important role in supergravity theory, known as the super-partner of the graviton, the gravitino \cite{sg}.  

Electrons and photons are particles accompanying us on daily life while the  RS particles are not found even in experiments of  high energy physics or in cosmology observations.  Recently, the interplaying between high energy physics and condensed matter physics supplies a new playground to the relativistic systems, e.g., Dirac semimetal in graphene \cite{graphene} and three dimensions \cite{dirac1,dirac2,dirac3},  topological insulator \cite{ti,ti1}, supersymmetric systems \cite{lee,yy,gdv,yao}, and newly proposed \cite{weyl} and discovered \cite{weyl1,weyl2,weyl4,weyl5,weyl6,weyl7,weyl8}  Weyl semimetal. To our knowledge, the RS physics was not concerned in the condensed matter  and cold atom context.  Can we expect  a RS or RS-Weyl semimetal in this playground?

The RS spinor can consist of the product of a spin-1 vector and a spin-1/2 Dirac spinor. The product can be decomposed into a spin-3/2  and a spin-1/2 irreducible representations of Lorentz group. After projecting to the spin-3/2 representation, this is the RS theory. The massless relativistic systems with spin$>0$ possess only two physical degrees of freedom, the highest and lowest  helicity states \cite{lurie}. Maxwell field with only transverse components is a well-known example.  There are also only two independent solutions of the massless RS equations,  the right- and left-handed chiral modes associated with helicity $\pm3/2$ .  This is a result of the Lorentz invariance and gauge invariance of the massless relativistic theory with spin$>1/2$.

Emergent Wely fermions in condensed matter systems were first recognized in the fermionic spectrum of
superfluid 3He-A and also in the core of quantized vortices in 3He-B \cite{vol}. The earlier theoretical prediction for Weyl semimetal was based on the $J_{eff}=1/2$ states in iridates \cite{weyl} which belong to a large class of 4$d$ and 5$d$ transition metal oxides with a strong spin-orbital coupling.  When the $d$-orbitals are below 2/3 filling, the crystal field projects the electron configurations to the $t_{2g}$ orbitals. When the strong spin-orbital coupling dominates, the electrons fill the $J_{eff}=1/2$ or $J_{eff}=3/2$ band depending on the filling factor. For iridates, the 5$d^5$ electrons occupy all $J_{eff}=3/2$ states and half fill $J_{eff}=1/2$ orbitals \cite{rev}.  For $d^{1,2}$ electrons, they quarterly or half occupy the $J_{eff}=3/2$ states. The representatives of these $J_{eff}=3/2$ materials are the ordered double perovskites with the chemical formula $A_2B'BO_6$ where $B'$ ions are commonly 4$d^{1,2}$ and 5$d^{1,2}$ transition metals', e.g. Mo$^{+5}$,Re$^{+6}$,Os$^{+7} $ for $d^1$ and  Re$^{+5}$,Os$^{+6}$ for $d^2$. Several exotic magnetic phases and a quadrupolar phase were presented in these strongly correlated materials\cite{rev,gchen1,gchen2}.
 This kind of  $J_{eff}=3/2$ systems may also exist  in p-orbital electrons of antiperovskites materials A$_3$BX \cite{capb} as well as cold atom systems  \cite{pwave}.  

We do not yet know in what kind of materials or cold atom systems the RS physics can exist. In this paper, we  present a recipe to realize the RS-Weyl semimetal from three orbital  Weyl semimetals with a strong on-site spin-orbital coupling. Each copy of Weyl semimetals is time reversal symmetry (TRS) breaking and possesses helicity, say $1/2$ at a right-handed Weyl point. We consider a large spin-orbital coupling limit of our model in order to understand the RS physics included in this system. In this limit, the on-site spin-orbital coupling  is taken as an unperturbed Hamiltonian and the zero order states are the eigen states of the total on-site angular momentum ${\bf J}$. There is a large energy gap between the $J=3/2$ and $1/2$ states. Projecting the three copies of Weyl semimetals to $J=1/2$, we  have a single copy of the Weyl semimetal with opposite helicity. Projecting to $J=3/2$,  there are two pairs of linear dispersions with different Fermi velocities.  The steeper dispersion is $|{\bf P}|$ while the flat dispersion is $|{\bf P}|/3$. Their helicities are 3/2 and 1/2 and yield the RS-Weyl and Weyl excitations, respectively.  The Berry phases of  helicity 3/2 and 1/2 states posses  the topological monopoles with charges $C=3$ and $C=1$, respectively, for an original $C=1$ Weyl point. To split the degeneracy between the RS-Weyl degrees of freedom with helicity 3/2 and the Weyl's with helicity 1/2, a non-local potential is needed. {\it A RS-Weyl semimetal emerges in the helicity 3/2 band}. 
Two generalizations are direct: we can apply this recipe to double Weyl fermions \cite{double,doubleexp,jianyao} and $^3$He-A with a triplet $p$ wave pairing \cite{vol,vol1}. 
We also study a RS-Weyl semimetal model with the space inversion symmetry (SIS) breaking. Projecting to $J_{eff}$= 3/2, we find the  same projected model as that in the TRS breaking systems.

This paper was organized as follows: In Sec. II, we will give a recipe for RS-Weyl semimetal with TRS breaking; In Sec. III, the generalizations and similarity with SIS mentioned in last paragraph are studied. The fourth section is our conclusions and discussions.

\section{Recipe for $J_{eff}=3/2$ semimetal with TRS breaking}

In this section, we study RS-Weyl fermion with TRS breaking.

\subsection{ Rarita-Schwinger equations} 

We  briefly introduce the RS equations. The RS equations for a sixteen component vector-spinor field $\psi_{\mu\alpha}$ in 3+1 dimensions are given by
\begin{eqnarray}    
&&(i\gamma^\mu\partial_\mu-m)\psi_\nu=0,\label{rs1}\\
&&\chi=\gamma^\mu\psi_\mu=0,\label{rs2}
\end{eqnarray} 
where the convention we use are: $\mu=0,a$  ($a=1,2,3$) denote the time-space indices with flat metric $\eta^{00}=\eta_{00}=1$ and $\eta_{aa}=-\eta^{aa}=-1$; $\alpha=s\sigma$ for $s=R,L$ and $\sigma=\uparrow,\downarrow$ are chiral and spin indices, respectively.  $\gamma^\mu$ are the gamma matrices.
In Eq. (\ref{rs2}), the four vector indices are  contracted over so that $\chi$ is a pure Dirac spinor. $\chi=0$ projects out the spin-1/2 sector and leaves only the degrees of freedom of the spin-3/2 sector.  It was known that if $m\ne 0$, there will be  fermonic modes with superluminal velocities if the RS field couples to the external electromagnetic field in a minimal way \cite{mass1,mass2}. 
A massless RS theory is gauge invariant under $\psi_\mu\to \psi_\mu+\partial_\mu\epsilon$ for an arbitrary spinor $\epsilon$. The gauge invariance of massless RS theory allowed us to take $\psi_0=0$ gauge as taking $A_0=0$ gauge for the electromagnetic field. Also similar to that in the Maxwell theory, only the transverse fields are physical degrees of freedom, namely,  $\partial_i\psi^i=0$. These transverse spinor fields are the
  right-(left-)handed fields with helicity 3/2 (-3/2) \cite{lurie,free}. We call them the right-(left-)handed RS-Weyl fields $ c_{aR(L)\sigma}$ if we write the four-component spinor $\psi^\dag_a=(c^\dag_{aL\sigma},c^\dag_{aR\sigma})$. Fourier modes of $\psi_a$ are denoted as $(c_{\pm3/2,p}{\cal U}_{aR(L)\sigma},d^\dag_{\pm3/2,p}{\cal V}_{aR(L)})$
 where  $c_{\pm3/2,p}$ and $d^\dag_{\pm3/2,p}$ are the particle and antiparticle modes with helicity $\pm3/2$.  ${\cal U}_{R(L)a}$ and ${\cal V}_{R(L)a}$ are two component spinors for a given $a$; They are normalized by
  ${\cal U}^{a\dag}_L\sigma^b{\cal U}_{aR}={\cal V}^{a\dag}_L\sigma^b{\cal V}_{aR}=-p^b/p_0$ and ${\cal U}^{a\dag}_{R(L)}{\cal V}_{aR(L)}=0$.  Notice that ${\cal U}$ and ${\cal V}$ are not independent but related by the charge conjugation. 
 The dispersions of these traverse modes are linear, i.e., $E=p_0=|\bf p|$, as expected \cite{lurie}.

\subsection{ Recipe for $J_{eff}=3/2$ semimetal with TRS breaking}
  
We now dispense a recipe for the RS-Weyl semimetal from a three orbital Weyl semimetal, with help of a strong on-site spin-orbital coupling. The Hamiltonian describes three copies of a Weyl semimetal on three-dimensional lattice, i.e.,
\begin{eqnarray}
H=\sum_{abc\sigma\sigma'\bf p}c^{\dag}_{a{\bf p}\sigma}[P_c({\bf p})\delta_{ab}\sigma_{c\sigma\sigma'}-\lambda L^c_{ab}\sigma_{c\sigma\sigma'}]c_{b\sigma'\bf p},\label{ham}
\end{eqnarray}
where the on-site orbital angular momentum matrix is defined by $L^1=T^1, L^2=-T^2$ and $L^3=-T^3$ with $T^c_{ab}=-i\epsilon_{abc}$; The second term  in (\ref{ham}) is on-site spin orbital coupling. We here assume $\lambda>0$. For any given $a$, ${\bf P}({\bf p})\cdot\boldsymbol{\sigma}$ in (\ref{ham}) describes a spin-1/2 Weyl semimetal with the TRS breaking in a condensed matter or cold atom system \cite{weyl}. Namely, near any right- or left-handed chiral Weyl point ${\bf p}_w$, ${\bf P}\cdot \boldsymbol{\sigma}\approx \pm v_F({\bf p}-{\bf p}_w)\cdot\boldsymbol{\sigma}$. 
For example, $P^1=t\sin p_1,P^2=t\sin p_2,P^3=2t(\cos p_3-\cos p_0)+m(2-\cos p_1-\cos p_2)$, which describe the Weyl semimetal in iridates A$_2$Ir$_2$O$_7$ \cite{exam}.
Eq. (\ref{ham}) is a single particle Hamiltonian and can be diagonalized. The dispersions of six branches are given by
\begin{eqnarray}
E&=&\pm|{\bf P}|-\lambda, \label{1} \\
E&=&\frac{1}2[\pm\sqrt{9\lambda^2+4({\bf P}^2\pm\lambda|{\bf P}|)}+\lambda]. \label{2}
\end{eqnarray}
The spin-orbital coupling lifts the degeneracy of three identical Weyl semimetals while
the Weyl points are the same as those in a single copy of Weyl semimetals. 
For a vanishing spin-orbital coupling, i.e., $\lambda<1/L$ where $L$ is the system size, (\ref{2}) reduces to $\pm|{\bf P}|+O(1/L)$.  Each copy of semimetals contributes a monopole charge $C=1$ of the Berry phase of the wave function surrounding a right-handed chirality Weyl point. However, the vanishing spin-orbital coupling is an isolate point. For any finite $\lambda$, i.e., $\lambda>1/L$,  the energy of two branches of (\ref{2}) near Weyl points lifts $2\lambda$ while other two's lowers $-\lambda$, the same as that in (\ref{1}). We calculate the  monopole charges by using $P^a$ in [\onlinecite{exam}]. For a right-handed Weyl point $(0,0,p_0)$,  instead of $C=1$ when $p_z\in (-p_0,p_0)$,
 the charge corresponding to (\ref{1})  becomes $C=3$. A $C=\pm1 $ monopole-anti-monopole pair develops, corresponding to (\ref{2}). The anti-monopole has a higher energy $2\lambda$ while the monopole with $C=1$ has the same energy as that with $C=3$.
If $\lambda$ is larger enough, say the order of  the bandwidth, the $C=1,3$ branches and $C=-1$ branches are separated into two bands: a lower band and a upper band, respectively. 

The monopole charges of a Weyl point for a finite $\lambda$ can be determined in the large $\lambda$ limit, which is not model-dependent. One can also see the RS degrees of freedom in this limit. We take the spin-orbital coupling term as the unperturbed Hamiltonian, denoted as $H_0$ and the Weyl semimetal part  in (\ref{ham}) as the perturbed Hamiltonian $H_1$. 
The total on-site angular momentum matrix  ${\bf J}_{a\sigma,b\sigma'}={\bf L}_{ab}\delta_{\sigma\sigma'}+\frac{1}2\delta_{ab}\boldsymbol{\sigma}_{\sigma\sigma'}$ commutes with the unperturbed Hamiltonian matrix ${\cal H}_{0}=-\lambda L^c_{ab}\sigma_{c\sigma\sigma'}$. The unperturbed wave functions are then the eigen states of $|{\bf J}|$. The eigen values of $|{\bf J}|$ are $J=1/2$ and 3/2 and the basis of unperturbed state spaces is given by
\begin{eqnarray}
\left[\begin{array}{l}
|\frac{1}2,\frac{1}2\rangle \\ 
|\frac{1}2,\frac{-1}2\rangle\\
|\frac{3}2,\frac{3}2\rangle\\
|\frac{3}2,-\frac{3}2\rangle\\
|\frac{3}2,\frac{1}2\rangle\\
|\frac{3}2,-\frac{1}2\rangle\end{array}\right]=\left[\begin{array}{ccccccc}
0&\frac{1}{\sqrt3}&0&\frac{i}{\sqrt3}&\frac{1}{\sqrt3}&0\\ 
\frac{1}{\sqrt3}&0&\frac{-i}{\sqrt3}&0&0&\frac{1}{\sqrt3}\\
\frac{-1}{\sqrt2}&0&\frac{-i}{\sqrt2}&0&0&0\\
0&\frac{1}{\sqrt2}&0&\frac{-i}{\sqrt2}&0&0\\
0&\frac{-1}{\sqrt6}&0&\frac{-i}{\sqrt6}&\sqrt{\frac{2}3}&0\\
\frac{1}{\sqrt6}&0&\frac{-i}{\sqrt6}&0&0&\sqrt{\frac{2}3}
\end{array}\right]
\left[\begin{array}{c}
|1,\uparrow\rangle\\
|1,\downarrow\rangle\\
|2,\uparrow\rangle\\
|2,\downarrow\rangle\\
|3,\uparrow\rangle\\
|3,\downarrow\rangle\end{array}\right]
\end{eqnarray}

The unperturbed energies are $2\lambda$ and $-\lambda$ corresponding to $J=1/2$ and $J=3/2$, respectively. In the large $\lambda$ limit, there is a large energy gap $\sim 3\lambda$ between the $J=1/2$ upper band and the $J=3/2$ lower band.  The projected matrices 
 of any $6\times6$ matrix ${\cal O}$ is defined by 
\begin{eqnarray}
{\cal O}_{1/2}={\cal P}^\dag_{1/2}{\cal O}{\cal P}_{1/2},
{\cal O}_{3/2}={\cal P}^\dag_{3/2}{\cal O}{\cal P}_{ 3/2},
\end{eqnarray}
 where the project matrices are given by
 \begin{eqnarray}
{\cal P}_{1/2}=\left[\begin{array}{ccc}
0&\frac{1}{\sqrt3}\\
\frac{1}{\sqrt3}&0\\
0&\frac{i}{\sqrt3}\\
\frac{-i}{\sqrt3}&0\\
\frac{1}{\sqrt3}&0\\
0&\frac{1}{\sqrt3}\end{array}\right],
{\cal P}_{3/2}=\left[\begin{array}{cccc}
\frac{-1}{\sqrt2}&0&0&\frac{1}{\sqrt6}\\
0&\frac{1}{\sqrt2}&\frac{-1}{\sqrt6}&0\\
\frac{i}{\sqrt2}&0&0&\frac{i}{\sqrt6}\\
0&\frac{i}{\sqrt2}&\frac{i}{\sqrt6}&0\\
0&0&\sqrt{\frac{2}3}&0\\
0&0&0&\sqrt{\frac{2}3}\end{array}\right]
\end{eqnarray}
Projecting to $J_{eff}=1/2$, we have an effective Weyl semimetal described by the following Hamiltonian 
 \begin{eqnarray}
 {\cal H}_{1/2}=\frac{1}3\left[\begin{array}{cccc}
-P_3&P_-\\
P_+&P_3\\
\end{array}\right]. \label{ph}\end{eqnarray}
It is a single copy of the Weyl semimetal with opposite helicity to the original Weyl semimetal. If we neglect the second order correction $O(|\bf P|^2/\lambda)$, the spectrum of the $J_{eff}=1/2$ system is simply $$E_{1/2,\pm1/2}= \pm|{\bf P}|/3+2\lambda.$$

Projecting to $J_{eff}=3/2$, one finds that ${\bf L}_{3/2}={\cal P}^\dag_{3/2}({\bf L}_{ab}\delta_{\sigma\sigma'}){\cal P}_{3/2}$
obey commutation relations 
$$[\frac{3}2L_{3/2}^a,\frac{3}2L_{3/2}^b]= \sum_{c=1}^3i\epsilon^{abc}\frac{3}2L_{3/2}^c.$$
That is, $J^a_{eff}=\frac{3}2 L^a_{3/2}$ are the SU(2) $J_{eff}=3/2$ generators. 
 In fact, the projected orbital and spin matrices are equal 
\begin{eqnarray}
{\cal P}^\dag_{3/2}({\bf L}_{ab}\delta_{\sigma\sigma'}){\cal P}_{3/2}={\cal P}^\dag_{3/2}(\delta_{ab}\boldsymbol {\sigma}_{\sigma\sigma'}){\cal P}_{3/2}. \label{magic}
\end{eqnarray}
Namely, the projected effective total on-site angular momentum matrix is given by 
\begin{eqnarray}
{\bf J}_{3/2}&=& {\cal P}^\dag_{3/2}({\bf L}_{ab}\delta_{\sigma\sigma'}){\cal P}_{3/2}+\frac{1}2{\cal P}^\dag_{3/2}(\delta_{ab}\boldsymbol{\sigma}_{\sigma\sigma'}){\cal P}_{3/2}\nonumber\\&=&\frac{3}2 {\cal P}^\dag_{3/2}({\bf L}_{ab}\delta_{\sigma\sigma'}){\cal P}_{3/2}=\frac{3}2{\bf L}_{3/2}.
\end{eqnarray}
This exactly gives rise to ${\bf J}_{3/2}={\bf J}_{eff}$.
And the projected Hamiltonian $
{\cal H}_{3/2}={\cal P}^\dag_{3/2}{\cal H}_1{\cal P}_{ 3/2}
$
reads
\begin{eqnarray}
\hspace{-2mm}{\cal H}_{3/2}={\bf P}\cdot {\bf L}_{3/2}=\frac{2}3{\bf P}\cdot {\bf J}_{3/2}=\left[\begin{array}{cccc}
P_3&0&\frac{P_-}{\sqrt3}&0\\
0&-P_3&0&\frac{P_+}{\sqrt3}\\
\frac{P_+}{\sqrt3}&0&\frac{P_3}3&\frac{2P_-}{3}\\
0&\frac{P_-}{\sqrt3}&\frac{2P_+}{3}&-\frac{P_3}3
\end{array}\right]~\end{eqnarray}
where $P_\pm=P_1\pm iP_2$.

Neglected the higher order correction, the dispersions of ${\cal H}_{3/2}+{\cal H}_0$ are given by
 \begin{eqnarray}
E_{3/2,\pm3/2}&=&\pm|{\bf P}|-\lambda,\label{pm32}\\
E_{3/2,\pm1/2}&=&\pm |{\bf P}|/3-\lambda.\label{pm12}
\end{eqnarray}
We see that the Weyl points  are not shifted. 

\subsection{ Helicity and RS-Weyl semimetal}

 The large $\lambda$ perturbed dispersions $E_{1/2,\pm1/2}$, $E_{3/2,\pm3/2}$ and $E_{3/2,\pm1/2}$ are of course consistent with the exact results (\ref{1}) and (\ref{2}). However, the simple Hamiltonians in the large $\lambda$ limit may explicitly give rise to more informations. The gapless linear dispersions imply that the eigen states are not the eigen states of the ${\bf J}_{eff}$. They are the eigen states of helicity operator. For $J_{eff}=1/2$, the helicity  operator is the same as the usual Weyl semimetal. Here we consider the case of $J_{eff}=3/2$. We define an operator $\mathfrak{h}=\hat{\bf P}\cdot {\bf J}$ with 
$\hat {\bf P}=\frac{{\bf P}}{|{\bf P}|}$. Projecting to $J_{eff}=3/2$,  the projected helicity operator is $\mathfrak{h}_{3/2}={\cal P}_{3/2}\hat{\bf P}\cdot {\bf J}{\cal P}_{3/2}=\hat{\bf P}\cdot {\bf J}_{3/2}$, i.e., the projected helicity operator matrix is given by
\begin{eqnarray}
h_{3/2}=\frac{3}{2|{\bf P}|}{\cal H}_{3/2}.
\end{eqnarray}
 This means that the projected helicity operator commutes with the Hamiltonian $H_{3/2}$. The states with the dispersion $E=\pm|\bf P|$ is the helicity eigen state with  $h_{3/2}=\pm3/2$ while the states with $E=\pm|{\bf P}|/3$ have the helicity $\pm1/2$. The corresponding monopole charges then are $C=3$ and $C=1$ for the right-handed Weyl point, recovered our calculation before. The total monopole charge in the $J_{eff}=3/2$ band is $C=4$.

The helicity $\pm3/2$ states give the RS-Weyl degrees of freedom in the  $J_{eff}=3/2$ band. Since $C=4$ is a topological invariant, any local perturbation can not split the helicity 1/2-sector  from the 3/2-sector in the lower band. One can add a non-local potential to lift this degenerate, e.g., 
\begin{eqnarray}
V=U\sum_{{\bf p},ij}c^\dag_{\bf p}h^2_{3/2}({\bf p})c_{\bf p} \propto \sum_{{\bf p},ij}\frac{P_aP_b}{|{\bf P}|^2}c^\dag_{i\bf p}[J_{3/2}^aJ_{3/2}^b]_{ij}c_{j\bf p}\label{nlp},\nonumber
\end{eqnarray}
where $U$ is a constant; $i$ and $j$ label four states with $J_{eff}^z=\pm1/2,\pm3/2$.  
This potential lifts the energy $E_{3/2,\pm3/2}\to E_{3/2,\pm3/2}+9U/4$ and $E_{3/2,\pm1/2}\to E_{3/2,\pm1/2}+U/4$.  Therefore, if $\lambda> |U|>$ bandwidth, the lower band is separated into a $h_{3/2}=\pm1/2$ sub-band and a  $h_{3/2}=\pm3/2$ sub-band. When Fermi energy is in the band of the helicity $\pm3/2$, the low-lying excitations near each Weyl point can be thought as the particle,say $c^\dag_{3/2,\bf p}$ and antiparticle $d^\dag_{3/2,\bf p}$, i.e., an emergent RS-Weyl semimetal. The physical origin of the non-local potential (\ref{nlp}) needs to be further studied. It may come from a long range interaction. 

In sum, we have offered a recipe for a RS semimetal. The candidates of possible materials are 
the condensed matter or cold atom systems in which the on-site spin and orbital degrees of freedom are strongly coupled. 

\section{Generalizations to Weyl fermions with SIS breaking, the multi-degeneracy, and paired wave function}

We now study the generalizations of our recipe to other systems: The Weyl fermions with SIS breaking, the multi-degeneracy, and paired wave function.

\subsection{ With SIS breaking}

The Weyl semimetal breaks either TRS or SIS. 
We can also start from a SIS breaking Weyl semimetal. 
We study the following Hamiltonian 
\begin{eqnarray}
H'_1=\sum_{abc\sigma{\bf p}}c^\dag_{a\sigma{\bf p}}L^c_{ab}P_cc_{b\sigma{\bf p}}
\end{eqnarray} 
Instead of Pauli matrices in $H_1$ which breaks the TRS, ${\bf L}$ connects the different orbital degrees of freedom in $H'_1$ and then breaks SIS.  The model $H_0+H_1'$ can also be analytically solved. The spectra read
\begin{eqnarray}
E&=&\pm|{\bf P}|-\lambda,\\
E&=&\frac{1}2[\lambda+|{\bf P}|\pm\sqrt{9\lambda^2+2\lambda|{\bf P}|+{\bf P}^2}],\\
E&=&\frac{1}2[\lambda-|{\bf P}|\pm\sqrt{9\lambda^2-2\lambda|{\bf P}|+{\bf P}^2}].
\end{eqnarray}
The $\lambda=0$ is also an isolated point. We are interested in finite $\lambda$ as well. In the strong coupling limit, one has the dispersions $E=\pm|{\bf P}|-\lambda$ with $C=3$,   $E=\pm|{\bf P}|/3-\lambda$ with $C=1$ and $E=\pm2|{\bf P}|/3+2\lambda$ with $C=-1$. 
The physics in each projected band is  the same as the cases with TRS breaking.

While there already are several models with TRS breaking \cite{weyl,exam}, we here give a toy model for $H_1'+H_0$ in a cubic lattice. The possible physical systems are cold atom gas with $p$-orbital coupling to pseudospin or $4d^1$ and $5d^1$ electrons with $t_{2g}$ orbital degrees of freedom.  
The Hamiltonian we are studying is as follows    
\begin{eqnarray}
H&=&-t\sum_{i\sigma} (\pm c_{2i\sigma}^\dag c_{3,i\pm\delta_x\sigma}\pm c^\dag_{1i\sigma}c_{2,i\pm\delta_3\sigma}\nonumber\\
&\pm& c^\dag_{1i}c_{3,i\pm\delta_y\sigma})+h.c.+H_0\label{lh}
\end{eqnarray}
where $\delta_{x,y,z}$ are the lattice vectors in the positive directions. The hopping term is between the different orbitals in different direction. Furthermore, the hopping in the negative $\delta_i$ direction carries a phase $\pi$ while carrying no phase in the positive direction. 
  After Fourier transformation, we have 
 $$P_a=t\sin p_a.$$ 
The Weyl points are located in those high symmetric points $\{(0,0,0),(\pi,\pi,\pi)\}, \{(0,\pi,\pi),(\pi,0,0)\},$ $\{(\pi,0,\pi),(0,\pi,0)\}$ and $\{(\pi,0,0),(0,0,\pi)\}$.  At these Weyl points, the effective Hamiltonian in the long wave length limit is exactly gravitino Hamiltonian in supergravity theory \cite{sg}.

\begin{figure}
\begin{center}
\includegraphics[width=7.5cm]{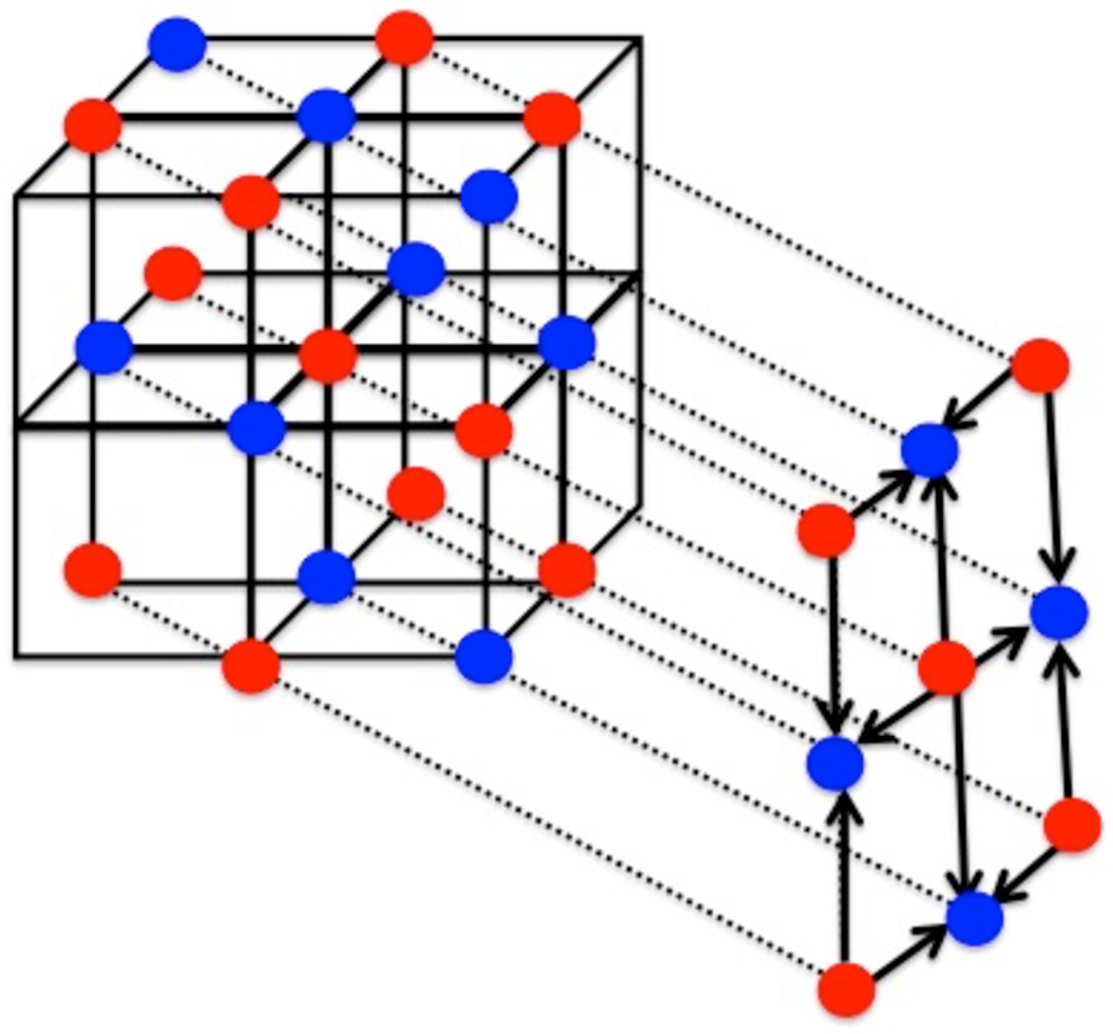}
\includegraphics[width=5.5cm]{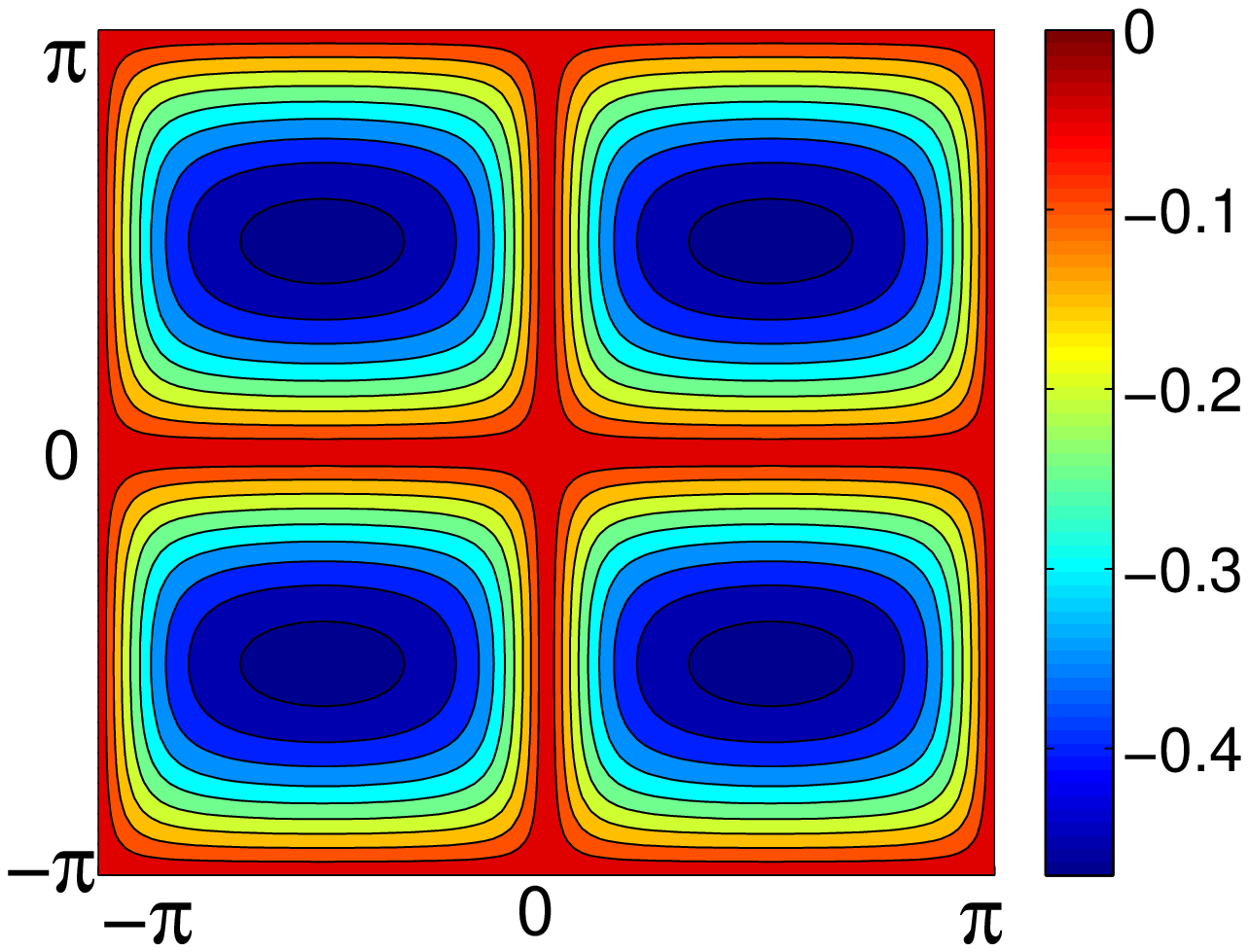}
\end{center}
 \caption{\label{fig3} (color online) The dispersions and the  Fermi arcs  in the [110]-surface Brillouin zone for $t=1$.Upper: Images of the Weyl points with charges $\pm 1$ (red and blue). Lower: The dispersions and the network of gapless Fermi arcs. }
\end{figure}  
\vspace{2mm}

We study the surface states of this model and consider a surface Brillouin zone with the normal vector along  the $(n_1,n_2,n_3)$ direction. The bulk Weyl points are projected onto this surface. A set of the Weyl points may have the same image. The charge of the image is defined by the sum of the winding numbers of these Weyl points in this set divided by the number of the points.  It is easy to see the charges of all images are zero if $n_1+n_2+n_3$ is odd while the charges of the images are $\pm 1$ if $n_1+n_2+n_3$ is even.  For a given surface, these images form a lattice. For a lattice where all sites are zero charged, there is no reason to see the Fermi arc connecting any pair of images while the Fermi arcs may appear between the positive and negative charged images in a charged site lattice.In Fig. \ref{fig3}(upper), we sketch the image network for an even surface, the [110] surface. The red and blue spots correspond to the images with charge $\pm 1$. 
 We numerically calculate the surface states of the [001], [111] and [110] surfaces. We see that there is no gapless state for [001] and [111] . The dispersions of the surface states for the [110] are shown in Fig. \ref{fig3}(lower). 
 As expected, the gapless Fermi arcs form a rectangular network and the images are the lattice sites of the network.

\subsection{Mutil-Weyl semimetals}

We consider a three-dimensional crystal that is invariant under an $n$-fold rotation about $z$-axis with
$n = 2, 3, 4, 6$. The Hamiltonian is also invariant under $C_m$ if $n/m$ is an integer.  If the tight-binding Hamiltonian $H({\bf p})={\bf P}({\bf p})\cdot\boldsymbol{\sigma}$ in (\ref{ham}) with translational symmetry \cite{double},
$C_m$-invariance gives
\begin{eqnarray} 
\hat C_m \hat H ({\bf p}) \hat C^{-1}_m = \hat H (R_m {\bf p}),
\end{eqnarray}
where $\hat C_m$ is the $m$-fold rotation operator and $R_m$ is the
$3\times 3$ rotation matrix. Eq. (\ref{ham}) defines three copies of multi-Weyl semimetal around a Weyl node through a strong spin-orbital coupling. After projecting to $J_{eff}=3/2$, we define a multi-RS-Weyl semimetal. For example, we consider a simple tight-binding model with $C_4$ symmetry given in [\onlinecite{jianyao}].  In a given Weyl point,  
the multi-Weyl semimetal Hamiltonian is given by
\begin{eqnarray}
\mathbf{P}\cdot\bm{\sigma}=(p_x^2-p_y^2)\sigma_x+2p_xp_y\sigma_y+p_z\sigma_z.
\end{eqnarray} 
In this case, the monopole charge is 2 instead of 1. Substituting into Eq. (\ref{ham}) and we see that with the help of strong spin-orbital coupling, three copies of double Weyl semimetal can also be projected to $J_{eff}=1/2$ with monopole charge -2 and $J_{eff}=3/2$ with  total monopole charge 8. For $J_{eff}=3/2$ states, the helicity 3/2 state has monopole charge 6 and helicity 1/2 state has monopole charge 2.  If we use the non-local potential (\ref{nlp}) in Sec. II C, we obtain a multi-RS-Weyl semimetal with monopole charge 6.

\subsection{Weyl fermions with paired wave function in $^3$He-A}

The $A$-phase of Helium 3 superfluid around a Weyl point can be described by the {\it B-dG} Hamiltonian \cite{vol1}
\begin{eqnarray} 
H=\left[\begin{array}{cccc}
p_z & 0 & -p_+& 0 \\ 
0 & p_z & 0 & p_+ \\ 
-p_-& 0 & -p_z & 0 \\ 
0 & p_- & 0 & -p_z
\end{array} \right].\label{b-dg}
\end{eqnarray}
There are two degenerate Weyl points with linear dispersion, and the total monopole charge is 2.  This is different from the case in the last subsection where the dispersion is quadratic in $x,y$ directions. If we couple three copies of this model by strong on-site spin-orbital coupling as before, the projected  effective Hamiltonian to  $J_{eff}=1/2$ states is given by
\begin{eqnarray} 
H_{1/2}=\frac{1}{3}\left[\begin{array}{cccc}
3p_z & 0 & -p_+ & 0 \\ 
0 & 3p_z & 0 & p_+ \\ 
-p_- & 0 & -3p_z & 0 \\ 
0 & p_- & 0 & -3p_z
\end{array} \right].
\end{eqnarray}
This has the same monopole charge as Eq. (\ref{b-dg}) , i.e., 2.

Projected to $J_{eff}=3/2$ states, the effective Hamilton is given by
\begin{eqnarray} 
H_{3/2}=\left[\begin{array}{cc}
p_z I_4&K\\
K^\dag&-p_z I_4
\end{array}\right]
\end{eqnarray}
where $I_4$ is $4\times 4$ unit matrix and 
\begin{eqnarray}
K=\left[\begin{array}{cccccccc}
 0 & 0 & 0 & p_+/\sqrt{3} \\ 
 0 & 0 & -p_+/\sqrt{3} & 0 \\ 
 0 & -p_+/\sqrt{3} & -2p_+/3 & 0 \\ 
 p_+/\sqrt{3} & 0 & 0 & 2p_+/3 
\end{array} \right]
\end{eqnarray}
This Hamiltonian is block-diagonal and describes RS fermions with pairing between (3/2,-1/2) and (-3/2,1/2). The eigen energies are 
still linear: $\pm |{\bf p}|$ and $\pm\frac{1}{3}\sqrt{9p_z^2+p^2_x+p^2_y}$ and the monopole charge is 4. Since the paired wave functions are no longer the eigenstates of the helicity, we can not separate these four dispersions by adding a non-local potential (\ref{nlp}).

  \section{Conclusions and discussions}

We have given a recipe for realizing the RS-Weyl semimetals from multi-copies of spin-1/2 Weyl fermion matters.  $J_{eff}=3/2$ chiral fermions are also of two components with helicity $\pm 3/2$. It was easy to be generalized to the multi-RS-Weyl and paired RS fermions and to obtain higher monopole charge Weyl points. 

We did not study  the  RS or RS-Weyl fermions in 2+1 dimensions. We make a simple discussions here and leave this to further study.   First, the massless RS equations in 2+1 dimensions are of only trivial solution. In 2+1 dimensions, the RS spinors $\psi_\mu$ ($\mu=0,1,2)$ are two components. and the gamma matrices consist of Pauli matrices:  $\gamma^0=\beta=\sigma^z$, $\gamma^1=\sigma^z\sigma^x=i\sigma^y$ and $\gamma^2=\sigma^z\sigma^y=-i\sigma^x$. Solving the supplementary conditions, one has
\begin{eqnarray}
\psi_0=-\sigma^x\psi_1-\sigma^y\psi_2, \psi_0=\frac{i}E(\partial_1\psi_1+\partial_2\psi_2).
\end{eqnarray} 
 Due to the gauge invariance, we can take $\psi_0=0$ gauge. Thus, if $\psi_a=\left(\begin{array}{c}\chi_a\\\xi_a\end{array}\right)$, then
 \begin{eqnarray}
&& \xi_2=i\xi_1, \chi_2=-i\chi_1;\nonumber\\
&& \partial_1\chi_1+\partial_2\chi_2=0,\partial_1\xi_1+\partial_2\xi_2=0 .
 \end{eqnarray}
Thus, $\xi_1=\xi_1(z)$ is holomorphic and $\chi_1=\chi_1(\bar z)$ are anti-holomorphic because
\begin{eqnarray}
\partial\chi_1=0, \bar\partial\xi_1=0.\nonumber
 \end{eqnarray}
 On the other hand, the Dirac equation for  $\psi_1$ gives
  \begin{eqnarray}
 E\chi_1(\bar z)=\partial\xi_1(z),E\xi_1(z)=\bar\partial\chi_1(\bar z).
 \end{eqnarray}
 And then $\psi_1=0$. Finally, we conclude that $\psi_\mu=0$ and there is only trivial solution of RS-Weyl equations. 
 
 The massive RS equations in 2+1 dimensions, especially its behavior in an external magnetic field are non-trivial \cite{mag}. We will study this in a separated work \cite{luoyu}.

\noindent{\bf Acknowledgements}   

The authors thank Gang Chen, Xi Luo, Yong-Shi Wu and Shiliang Zhu for helpful discussions.  This work is supported by the 973 program of MOST of China (2012CB821402), NNSF of China (11174298, 11474061). L.L. is partially supported by the Academy of Finland
through its Centres of Excellence Programme (2015-
2017) under project number 284621.

\end{document}